\documentclass[11pt]{article}
\usepackage{moriond,epsfig}

\def\be{\begin{equation}}
\def\ee{\end{equation}}
\def\bea{\begin{eqnarray}}
\def\eea{\end{eqnarray}}

\newcommand{\im}{\mathrm{Im\,}}

\newcommand{\mh}{m_h}
\newcommand{\z}{\frac{{m_h}^2}{4\epsilon_0^2}}

\begin{document}
\def\thefootnote{\fnsymbol{footnote}}
\setcounter{footnote}{1}
\title{Quarkonium hadron interaction in QCD~\footnote{Talk given at XXXVIth Rencontres de Moriond: QCD and High Energy Hadronic Interactions, Les Arcs, France, 17-24 March 2001.}
}
\author{F. Arleo}

\address{SUBATECH \\
Laboratoire de Physique Subatomique et des Technologies Associ\'ees\\
UMR Ecole des Mines de Nantes, IN2P3/CNRS, Universit\'e de Nantes\\
4, rue Alfred Kastler,
F-44070 Nantes Cedex 03, France.}

\maketitle\abstracts{
The analytic continuation of the operator product expansion of the scattering amplitude allows to compute the heavy-quarkonium hadron total cross section. The energy dependence of the $\Upsilon$ and $\Upsilon'$ cross sections with a proton is discussed.}
\setcounter{footnote}{0}

\section{Introduction}

Heavy quarkonium ($\Phi$) production is known to be sensitive to quark-gluon plasma formation in heavy ion collisions~\cite{mat}. The NA50 results on the anomalous $J/\psi$ suppression in Pb(158 AGeV)-Pb collisions~\cite{na50} triggered a lively and ongoing discussion on whether or not a ``new state of matter'' has been formed.

It has been conjectured that $J/\psi$ inelastic interactions with {\it comoving} particles produced in the collision may account for most of the $E_T$ dependence of the NA50 data~\cite{cap}. It proves therefore essential to determine the strength of the interaction of a heavy-quarkonium with a hadron.

In these proceedings, we shall first present a calculation of the $\Phi$-hadron cross section based on the operator product expansion~\cite{arl01}. Subsequently, the energy dependence of the $\Upsilon$ and $\Upsilon'$ cross sections will be discussed.

\section{$\Phi$ - $h$ scattering in QCD}

At leading twist, the scattering amplitude of a heavy quarkonium $\Phi$ with a target hadron $h$ reads~\cite{Pe}
\begin{equation}\label{eq:lt_amplitude}
\mathcal{M}_{\Phi\,h}(\lambda)=a_0^3\epsilon_0^2
\sum_{k\ge 1}d_{2k}\epsilon_0^{-2k}
\langle h| F^{0\nu}(iD^0)^{2k-2}F^{\phantom{\nu}0}_\nu |h\rangle,
\end{equation}
where $a_0$ and $\epsilon_0$ are respectively the Bohr radius and the Rydberg energy of the heavy-quark system. The LT amplitude in Eq.~(\ref{eq:lt_amplitude}) is an expansion of hard coefficients $d_{2k}$ and soft matrix elements $\langle h|\cdots|h\rangle$:

$\bullet\,$ The $d_{2k}$ coefficients are numbers (for spin averaged amplitude) that are matrix elements evaluated in the quarkonium state. These have been computed to leading order in QCD perturbation theory and in the large $N_c$ limit for 1S and 2S $\Phi$-states~\cite{Pe}. They may be expressed as the $2k$-th moments of a given function $f^{(nS)}$~\cite{arl01}
\[
d_{2k}^{(nS)} = \int_0^1 \frac{dx}{x} x^{2k} f^{(nS)}(x),
\]
with
\begin{eqnarray*}
f^{(1S)}(x)&=&\frac{16^3}{3N_c^2} x^{5/2}(1-x)^{3/2}\\
f^{(2S)}(x)&=&\frac{16^3}{3N_c^2} x^{5/2}(1-x)^{3/2}(2-6x)^2.
\end{eqnarray*}

$\bullet\,$ The matrix elements $\langle h|\cdots|h\rangle$ in Eq.(\ref{eq:lt_amplitude}) are expectation values of twist-2 gluon field operators in the incoming hadron $h$. They can be written as
\begin{equation}\label{eq:gluon_matrix_element}
\langle h(p)|F^{0\nu}(i:D^0)^{2k-2}F^{\phantom{\nu}0}_\nu|h(p)\rangle
=A_{2k}\sum_{j=0}^k \frac{(2k-j)!}{j!(2k-2j)!} 
(-m_h^2/4)^j \lambda^{2k-2j}
\end{equation}
where $\lambda$ is the hadron energy in the $\Phi$ rest frame and $A_{2k}$ is the Mellin transform of the gluon distribution $G^h$ in the hadron $h$, i.e.,
\[
A_{2k} = \int_0^1 \frac{dx}{x} x^{2k} G^h(x).
\]
We note that only the first term in Eq.~(\ref{eq:gluon_matrix_element}) will contribute to the soft matrix elements in the limit of massless hadrons. The inclusion of finite mass corrections ($j \ge 1$ terms in (\ref{eq:gluon_matrix_element})) on the cross sections has first been investigated by Kharzeev {\it al.}~\cite{KSSZ} and more recently in Ref.~\cite{arl01}.

~\\
From Eq.~(\ref{eq:lt_amplitude}) and Eq.~(\ref{eq:gluon_matrix_element}) and after a change of variable $k \to k'=k-j$, the forward scattering amplitude may be written as a double series
\begin{equation}\label{eq:power_series_p}
{\cal M}(\lambda,\mh)=a_0^3\epsilon_0^2
\sum_{j\ge 0,k'\ge 1}(\lambda/\epsilon_0)^{2k'}M_{2(k'+j)}
\frac{(2k'+j)!}{j!(2k')!}\left(-\z\right)^j
+a_0^3\epsilon_0^2\sum_{j\ge 1}M_{2j}
\left(-\z\right)^j
\end{equation}
where $M_{2k} = d_{2k}^{(nS)} A_{2k}$. The second term in the r.h.s. of Eq.~(\ref{eq:power_series_p}) that is real does not contribute to the total $\Phi$-$h$ cross section and will thus be dropped in the following. This (double) series is absolutely convergent provided $|\lambda| < \epsilon_0 - \mh^2/4\epsilon_0$, i.e. in the {\it unphysical} region of energies. One has therefore to continue analytically Eq.~(\ref{eq:power_series_p}) throughout the whole complex plane to determine the LT amplitude in the {\it physical} region of energies. This may be done first by noticing that $M_{2k}$ being the product of the $2k$-th moments of $f^{(nS)}$ and $G^h$ respectively may thus be written 
\begin{equation}\label{eq:moments}
M_{2k} = \int_0^1 \frac{dx}{x} \,x^{2k} f^{(nS)} \otimes G(x).
\end{equation}
Using Eq.~(\ref{eq:moments}) in Eq.~(\ref{eq:power_series_p}), swapping the sums with the integral and summing over $j$ and $k'$ leads eventually to the imaginary part of the LT amplitude~\footnote{Eq.~(\ref{eq:im_amplitude_p}) is exact provided the hadron mass $\mh < 2 \epsilon$. The calculation may similarly be carried out in the case $\mh \ge 2 \epsilon$ and can be found in Ref.~\cite{arl01}.}
\begin{equation}
\label{eq:im_amplitude_p}
\im\mathcal{M}(\lambda)=\frac{\pi}{2}a_0^3\epsilon_0^2\,
\frac{\lambda_+}{\sqrt{\lambda^2-\mh^2}}\,f^{(nS)} \otimes G(\epsilon_0/\lambda_+),
\end{equation}
where $\lambda_+=\Big(\lambda+\sqrt{\lambda^2-\mh^2}\Big)/2$ (see Ref.~\cite{arl01}). The amplitude (\ref{eq:im_amplitude_p}) is now well defined on the real axis. Dividing by the flux factor $\sqrt{\lambda^2 - \mh^2}$ leads to the total $\Phi$-$h$ cross section $\sigma_{\Phi h}$ {\it via} the optical theorem. Expliciting the product of convolution, we obtain
\begin{equation}\label{eq:cross_section_p}
\sigma_{\Phi(nS)\,h}(\lambda)=\frac{\lambda_+^2}{\lambda^2-\mh^2}
\int_0^1 dx\,G(x)\,\sigma_{\Phi(nS)\,g}(x\lambda_+).
\end{equation}
where $\sigma_{\Phi(nS)\,g}$ is defined as
\begin{eqnarray}\label{eq:cross_section_g}
\sigma_{\Phi(1S)\,g}(\omega)&=&\frac{16^3\pi}{6N_c^2}a_0^3\epsilon_0
\frac{(\omega/\epsilon-1)^{3/2}}{(\omega/\epsilon)^5}\theta(\omega-\epsilon)\\
\sigma_{\Phi(2S)\,g}(\omega)&=&16\frac{16^3\pi}{6N_c^2}a_0^3\epsilon_0
\frac{(\omega/\epsilon-1)^{3/2}(\omega/\epsilon-3)^2}
{(\omega/\epsilon)^7}\theta(\omega-\epsilon)
\end{eqnarray}
with the $\Phi(nS)$ state binding energy $\epsilon=\epsilon_0/n^2$.  Eq.~(\ref{eq:cross_section_p}) may have a simple partonic interpretation: as anticipated in Ref.~\cite{Pe}, the LT analysis describes the $\Phi$ dissociation by gluons from the hadron $h$~\cite{arl01}.

\section{Absolute cross sections}

As already mentioned, both the magnitude and the energy dependence of the heavy-quarkonium hadron cross sections prove to be of first importance in the context of heavy ion physics. Let us therefore address these very questions for the bottomonium channel in this section.

\begin{center}
\begin{figure}[htbp]
\begin{minipage}{7.5cm}
\centerline{\psfig{figure=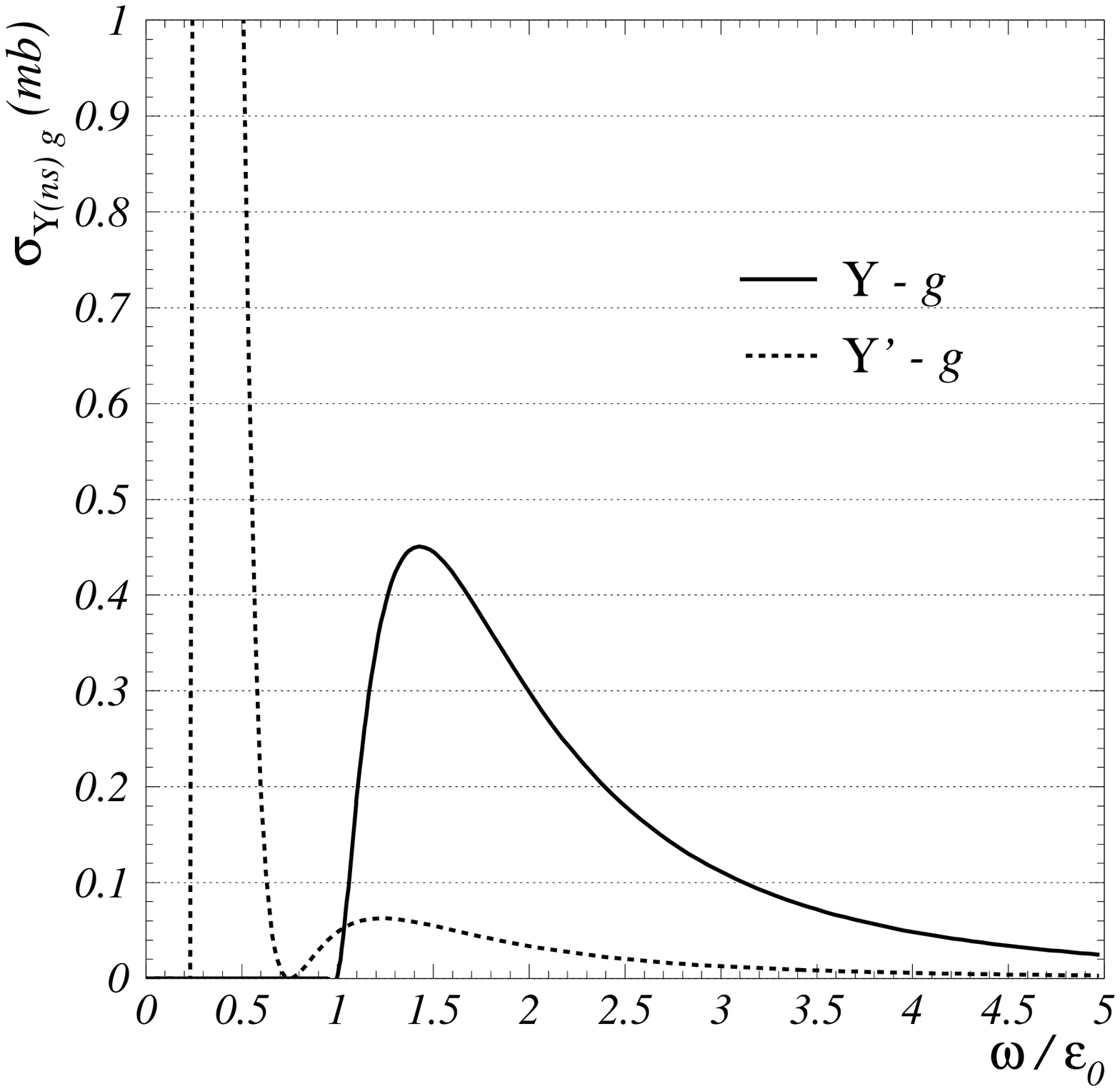,width=7.5cm}}
\end{minipage}
\hfill
\begin{minipage}{7.5cm}
\centerline{\psfig{figure=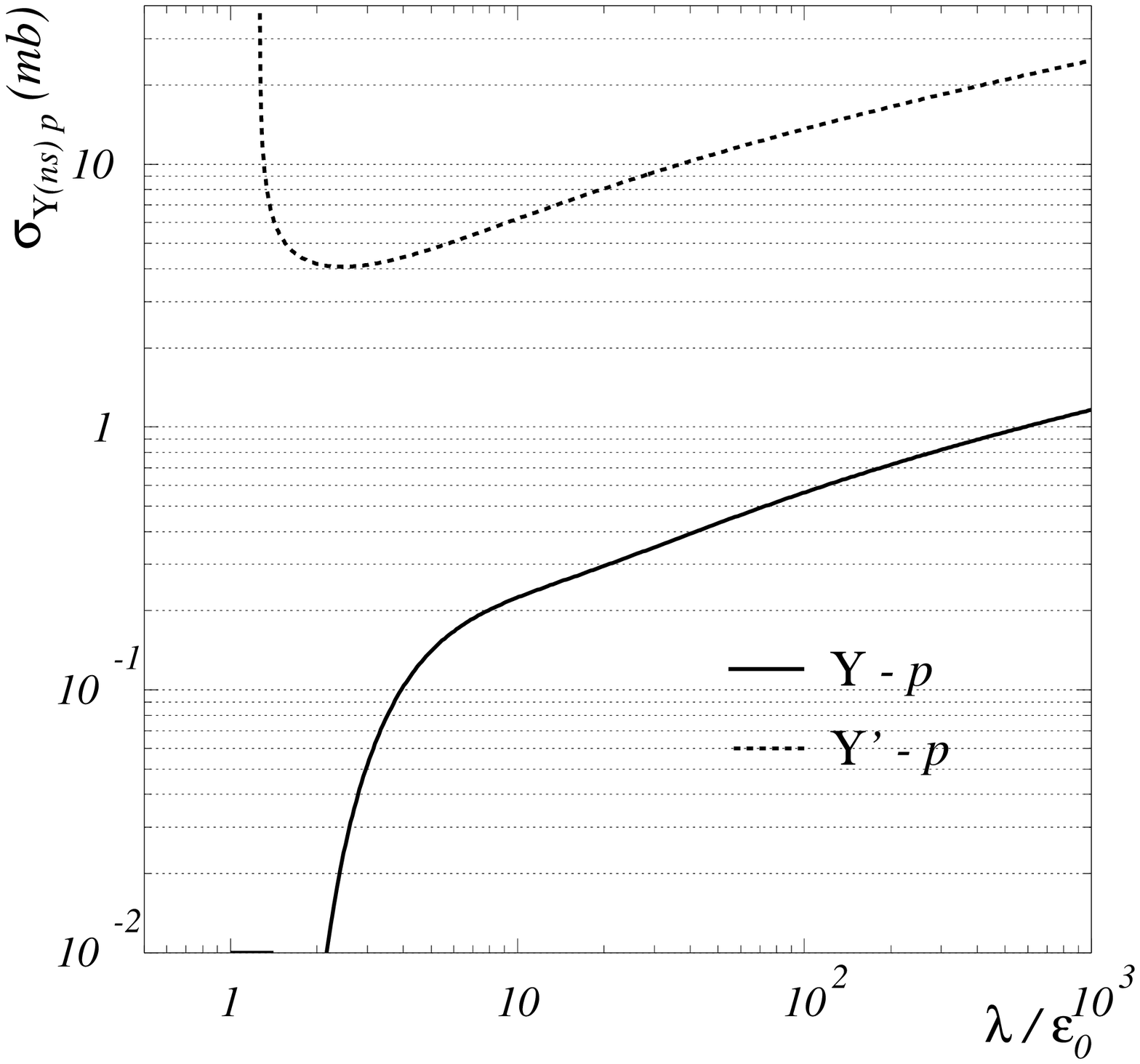,width=7.5cm}}
\end{minipage}
\caption{Energy dependence of the $\Upsilon$ ($solid$) and $\Upsilon'$ ($dashed$) cross section with a gluon ($left$) and a proton ($right$).}
\label{fig}
\end{figure}
\end{center}

The energy dependence of the partonic cross sections $\sigma_{\Phi(nS) g}$ is shown in Figure~\ref{fig} ($left$) for both $\Upsilon(1S)$ ($solid$) and $\Upsilon'(2S)$ ($dashed$) states~\footnote{The computations are performed here assuming a Bohr radius $a_0 =$~0.1 fm and a Rydberg energy $\epsilon_0 =$~0.75~GeV. See Ref.~\cite{arl01} for details.}. The cross sections are plotted as a function of $\omega / \epsilon_0$ with $\omega$ being the gluon energy in the bottomonium rest frame. Figure~\ref{fig} reveals that the partonic cross sections are sharply peaked for both 1S and 2S-states just above the dissociation threshold $\epsilon$ up to a maximum $\sigma_{\Upsilon g}^{\rm max} \approx 0.45$ mb and $\sigma_{\Upsilon' g}^{\rm max} \approx 12$ mb. Indeed, $\sigma_{\Phi(nS) g}$ is already 10\% of $\sigma_{\Upsilon(nS) g}^{\rm max}$ for gluon energies $\omega \ge 4-5\,\,\epsilon$. This characteristic has been attributed in Ref.~\cite{xu} to the small size (and thus small cross sections) probed by the high momentum gluons. Another feature is remarkable in Figure~\ref{fig} ($left$) : bottomonium may be probed by gluons carrying two {\it distinct} momenta whether the onium is $\Upsilon$ ($\omega\approx 1.4\,\epsilon_0$) or $\Upsilon'$ ($\omega\approx 1.2\,\epsilon = 0.3\,\epsilon_0$). Such an effect may be seen in the $p_T$ dependence of the $\Upsilon'/\Upsilon$ ratio. Using finite temperature potential calculations, Gunion and Vogt claimed that such an observable turns out to be a promising tool to study the quark-gluon plasma~\cite{gun}.

Figure~\ref{fig} ($right$) displays the cross sections of the $\Upsilon(nS)$ states with a proton as a function of $\lambda / \epsilon_0$. The GRV94~LO gluon distribution~\cite{GRV94} is used for the calculation. First, we notice that $\sigma_{\Upsilon p}$ rises dramatically just above the threshold located at $\lambda = \epsilon_0+\mh^2/ 4\epsilon_0$. Above $\lambda \sim 10\,\epsilon_0$, the $\Upsilon p$ cross section smoothly increases with the hadron energy, up to 1 mb at $\lambda / \epsilon_0 \sim 10^3$. As previously discussed, the $\Upsilon$ state is dissociated by gluons with energy $\sim \epsilon$. The smallness of this cross section at threshold is therefore simply understood as the need to find gluons with momentum fractions $x \sim \epsilon / \lambda \to 1$. At high energy $\lambda$, one may approximate Eq.~(\ref{eq:cross_section_g}) with $\sigma_{\Upsilon(nS) g}(\omega) \propto \delta(\omega/\epsilon-1)$, leading asymptotically to 
\[
\sigma_{\Upsilon(nS)\, p} \propto \frac{\epsilon}{\lambda}\,\, G^p\left(\frac{\epsilon}{\lambda}\right).
\]

The $\Upsilon'-p$ cross section is also shown ($dashed$) in Figure~\ref{fig}. Whereas the high energy behavior $\sigma_{\Upsilon' p}$ is found to be similar to the $\Upsilon$ channel, with an asymptotic ratio $\sigma_{\Upsilon' p} / \sigma_{\Upsilon p} \sim 20$, the energy dependence at low incident energy is opposite. At threshold, the $\Upsilon'$-p cross section becomes divergent. However, we do not expect any phenomenological implications whatsoever since the energy region for which the cross section gets large is small. Such a divergence occurs in the very case where the hadron mass $\mh$ is larger than twice the binding energy~$\epsilon$ (Ref.~\cite{arl01}).

To conclude this section, we emphasize that the energy dependence of $\sigma_{\Phi g}$ and $\sigma_{\Phi p}$ proves to be radically different. Whereas inelastic interactions of a heavy quarkonium with comoving {\it hadrons} seem to be unlikely, it appears that the nuclear dependence of heavy quarkonium production may strongly be affected by comoving {\it gluons}, as first pointed out in Ref.~\cite{ahm}.

\section{Summary}

Let us summarize what have been presented in these proceedings :

\begin{itemize}

\item First, the LT heavy-quarkonium hadron cross section has been given. The mass of the scattering hadron has systematically been incorporated in the cross section.

\item The proposed analysis has been performed for both 1S and 2S-states, although the small 2S-states binding energy does not fully justify a perturbative approach.

\item Finally, the energy dependence of both the $\Upsilon-p$ and $\Upsilon'-p$ cross sections has been investigated. The partonic $\sigma_{\Phi(nS) g}$ cross sections have also been discussed.

\end{itemize}

\section*{Acknowledgments}

This work has been done in collaboration with J.~Aichelin, P.-B.~Gossiaux, and T.~Gousset. I am also grateful to A. Capella and S. Peign\'e for discussions.


\section*{References}
\begin{thebibliography}{5}

\bibitem{mat} T.~Matsui and H.~Satz, Phys.\ Lett.\ {\bf B178}, 416
(1986).

\bibitem{na50} M. C. Abreu {\it et al.}, Phys.\ Lett.\ {\bf B477}, 28 (2000).\\
R. Arnaldi, these proceedings (hep-ex/0106079).

\bibitem{cap} N. Armesto, A. Capella, and E.G. Ferreiro, Phys.\ Rev.\ C {\bf 59}, 395 (1999).\\
D. Sousa, these proceedings (nucl-th/0106066).

\bibitem{arl01} F. Arleo, P.-B. Gossiaux, T. Gousset and J. Aichelin, hep-ph/0102095 (2001)

\bibitem{Pe} M.~E.~Peskin, Nucl.\ Phys.\ {\bf B156}, 356 (1979)\\
G.~Bhanot and M.~E.~Peskin, Nucl.\ Phys.\ {\bf B156}, 391 (1979).

\bibitem{KSSZ} D.~Kharzeev, H.~Satz, A.~Syamtomov, and G.~Zinovjev,
Phys.\ Lett.\ {\bf B389}, 595 (1996).

\bibitem{xu} X.~Xu, D.~Kharzeev, H.~Satz, and X.-N.~Wang, Phys.\ Rev.\ C {\bf 53}, 3051 (1996).

\bibitem{gun} J.~F.~Gunion and R.~Vogt, Nucl.\ Phys.\ {\bf B492}, 301 (1997). 

\bibitem{GRV94} M.~Gl\"uck, E.~Reya, and A.~Vogt, Z.\ Phys.\ C {\bf 67}, 433 (1995).

\bibitem{ahm} S.~J.~Brodsky and A.~H.~Mueller, Phys.\ Lett.\ {\bf B206}, 685 (1988). 

\end{thebibliography}
\end{document}